\documentclass[a4paper,USenglish,cleveref,autoref,thm-restate]{lipics-v2021}

\usepackage{amsmath}
\usepackage{mathtools}
\usepackage{xspace}
\usepackage{algorithm}
\usepackage{algpseudocode}
\usepackage{xcolor}
\usepackage{listings}

\newcommand{\tool}{\textsf{clone-finder}\xspace}

\newcommand{\cicApp}[2]{\ensuremath{#1~#2}}
\newcommand{\cicFun}[3]{\ensuremath{\mathtt{fun~(}#1~\mathtt{:}~#2\mathtt{)}~\Rightarrow~#3}}
\newcommand{\cicProd}[3]{\ensuremath{\mathtt{forall~(}#1~\mathtt{:}~#2\mathtt{),}~#3}}
\newcommand{\cicDef}[4]{\ensuremath{\mathtt{let}~#1~\mathtt{:}~#2~\mathtt{\coloneq}~#3~\mathtt{in}~#4}}
\newcommand{\cicMatch}[3]{\ensuremath{\mathtt{match}~#1~\mathtt{return}~#2~\mathtt{with}~#3~\mathtt{end}}}
\newcommand{\cicFix}[5]{\ensuremath{\mathtt{fix}~#1~\mathtt{(}#2~\mathtt{:}~#3\mathtt{)}~\mathtt{:}~#4~\mathtt{\coloneq}~#5}}
\newcommand{\cicSet}{\ensuremath{\mathtt{Set}}}
\newcommand{\cicType}{\ensuremath{\mathtt{Type}}}

\newcommand{\aEq}[2]{\ensuremath{#1=_\alpha #2}}
\newcommand{\alphaEquivalent}{$\alpha$-equivalent\xspace}
\newcommand{\alphaEquivalence}{$\alpha$-equivalence\xspace}
\newcommand{\rep}[3]{\ensuremath{#1^{#2\rightarrow #3}}}

\newcommand{\ie}{\textit{i.e.}\xspace}
\newcommand{\eg}{\textit{e.g.}\xspace}

\crefformat{section}{\S#2#1#3}
\crefformat{subsection}{\S#2#1#3}
\crefformat{subsubsection}{\S#2#1#3}

\newcommand\code[1]{{\tt\scriptsize #1}}
\definecolor{dkgreen}{rgb}{0,0.6,0}
\definecolor{ltblue}{rgb}{0,0.4,0.4}
\definecolor{dkblue}{rgb}{0,0,0.8}
\definecolor{dkred}{rgb}{0.8,0,0}
\definecolor{dkviolet}{rgb}{0.3,0,0.5}
\lstdefinelanguage{Coq}{ 
    mathescape=true,
    texcl=false, 
    morekeywords=[1]{Section, Module, End, Require, Import, Export,
        Variable, Variables, Parameter, Parameters, Axiom, Hypothesis,
        Hypotheses, Notation, Local, Tactic, Reserved, Scope, Open, Close,
        Bind, Delimit, Definition, Let, Ltac, Fixpoint, CoFixpoint, Add,
        Morphism, Relation, Implicit, Arguments, Unset, Contextual,
        Strict, Prenex, Implicits, Inductive, CoInductive, Record,
        Structure, Canonical, Coercion, Context, Class, Global, Instance,
        Program, Infix, Theorem, Lemma, Corollary, Proposition, Fact,
        Remark, Example, Proof, Goal, Save, Qed, Defined, Hint, Resolve,
        Rewrite, View, Search, Show, Print, Printing, All, Eval, Check,
        Projections, inside, outside, Def},
    morekeywords=[2]{forall, exists, exists2, fun, fix, cofix, struct,
        match, with, end, as, in, return, let, if, is, then, else, for, of,
        nosimpl, when},
    morekeywords=[3]{Type, Prop, Set, true, false, option},
    morekeywords=[4]{pose, set, move, case, elim, apply, clear, hnf,
        intro, intros, generalize, rename, pattern, after, destruct,
        induction, using, refine, inversion, injection, rewrite, congr,
        unlock, compute, ring, field, fourier, replace, fold, unfold,
        change, cutrewrite, simpl, have, suff, wlog, suffices, without,
        loss, nat_norm, assert, cut, trivial, revert, bool_congr, nat_congr,
        symmetry, transitivity, auto, split, left, right, autorewrite},
    morekeywords=[5]{by, done, exact, reflexivity, tauto, romega, omega,
        assumption, solve, contradiction, discriminate},
    morekeywords=[6]{do, last, first, try, idtac, repeat},
    morecomment=[s]{(*}{*)},
    showstringspaces=false,
    morestring=[b]",
    morestring=[d]’,
    tabsize=3,
    extendedchars=false,
    sensitive=true,
    breaklines=false,
    basicstyle=\scriptsize,
    captionpos=b,
    columns=[l]flexible,
    identifierstyle={\ttfamily\color{black}},
    keywordstyle=[1]{\ttfamily\color{dkviolet}},
    keywordstyle=[2]{\ttfamily\color{dkgreen}},
    keywordstyle=[3]{\ttfamily\color{ltblue}},
    keywordstyle=[4]{\ttfamily\color{dkblue}},
    keywordstyle=[5]{\ttfamily\color{dkred}},
    stringstyle=\ttfamily,
    commentstyle={\ttfamily\color{dkgreen}},
    literate=
    {\\forall}{{\color{dkgreen}{$\forall\;$}}}1
    {\\exists}{{$\exists\;$}}1
    {<-}{{$\leftarrow\;$}}1
    {=>}{{$\Rightarrow\;$}}1
    {==}{{\code{==}\;}}1
    {==>}{{\code{==>}\;}}1
    {->}{{$\rightarrow\;$}}1
    {<->}{{$\leftrightarrow\;$}}1
    {<==}{{$\leq\;$}}1
    {\#}{{$^\star$}}1 
    {\\o}{{$\circ\;$}}1 
    {\@}{{$\cdot$}}1 
    {\/\\}{{$\wedge\;$}}1
    {\\\/}{{$\vee\;$}}1
    {++}{{\code{++}}}1
    {~}{{$\sim$}}1
    {\@\@}{{$@$}}1
    {\\mapsto}{{$\mapsto\;$}}1
    {\\hline}{{\rule{\linewidth}{0.5pt}}}1
}[keywords,comments,strings]

\title{Automatic Goal Clone Detection in Rocq}

\titlerunning{Automatic Goal Clone Detection in Rocq}

\author{Ali Ghanbari}{Auburn University, United States \and \url{https://ali-ghanbari.github.io/} }{ghanbari@auburn.edu}{https://orcid.org/0000-0003-1471-2546}{}

\authorrunning{A. Ghanbari} 

\Copyright{Ali Ghanbari}

\end{CCSXML}



\supplement{}

\supplementdetails[subcategory={Source code}, cite={}, swhid={}]{Software}{https://aub.ie/cfsrc}

\nolinenumbers 

\EventEditors{Jonathan Aldrich and Alexandra Silva}
\EventNoEds{2}
\EventLongTitle{39th European Conference on Object-Oriented Programming (ECOOP 2025)}
\EventShortTitle{ECOOP 2025}
\EventAcronym{ECOOP}
\EventYear{2025}
\EventDate{June 30--July 2, 2025}
\EventLocation{Bergen, Norway}
\EventLogo{}
\SeriesVolume{333}
\ArticleNo{23}

\begin{document}

\maketitle

\begin{abstract}
Proof engineering in Rocq is a labor-intensive process, and as proof developments grow in size, redundancy and maintainability become challenges.
One such redundancy is \textit{goal cloning}, \ie, proving \alphaEquivalent goals multiple times, leading to wasted effort and bloated proof scripts.
In this paper, we introduce \tool, a novel technique for detecting goal clones in Rocq proofs.
By leveraging the formal notion of \alphaEquivalence for Gallina terms, \tool systematically identifies duplicated proof goals across large Rocq codebases.
We evaluate \tool on 40 real-world Rocq projects from the CoqGym dataset.
Our results reveal that each project contains an average of 27.73 instances of goal clone.
We observed that the clones can be categorized as either \textit{exact goal duplication}, \textit{generalization}, or \textit{\alphaEquivalent goals with different proofs}, each signifying varying levels duplicate effort.
Our findings highlight significant untapped potential for proof reuse in Rocq-based formal verification projects, paving the way for future improvements in automated proof engineering.
\end{abstract}

\section{Introduction}\label{sec:introduction}
Rocq~\cite{bib:coqDoc}\footnote{The Coq development community recently renamed the system to The Rocq Prover. In this paper, we use Coq and Rocq interchangeably.} is a popular interactive theorem prover based on type theory.
Rocq has proved to be a viable tool for constructing certified software, as demonstrated by projects like CompCert~\cite{bib:leroy2009formal}, Iris~\cite{bib:jung2018iris}, ConCert~\cite{bib:annenkov2020concert}, and Fiat Cryptography~\cite{bib:erbsen2020simple}.

\textit{Proof engineering}~\cite{bib:ringer2020qed} in Rocq is a labor-intensive task~\cite{bib:bourke2012challenges,bib:first2020tactok,bib:first2022diva}.
As proof developments grow in size, redundancy and maintainability become major challenges.
An important, yet relatively overlooked, issue in proof engineering research is \textit{goal cloning}--proving of \alphaEquivalent goals more than once.
Such a duplication of effort not only results in waste of manpower, but also inflates proof scripts, increases maintenance costs, and reduces opportunities for reuse.
While Rocq offers mechanisms for structuring proofs, avoiding redundancy remains a challenge.
Large-scale verification projects, such as seL4~\cite{bib:klein2010sel4} and Verisoft~\cite{bib:alkassar2009verisoft}, have highlighted the necessity of tools that support automated detection of redundant proof efforts~\cite{bib:bourke2012challenges}.
While there is tool support for code clone detection~\cite{bib:ghanouchi2023clone,bib:vinan2016clone} in Isabelle/HOL~\cite{bib:nipkow2002isabelle}, to the best of our knowledge, detection of redundant proof efforts in Rocq proofs is not explored yet.

The problem that we set out to solve is as follows.
\textit{Given a set of Rocq proofs, find the set of \alphaEquivalent goals making up the proofs.}
``Goal'' is a Rocq terminology for proof obligation; each proof comprises sub-proofs for a set of goals that are written as Gallina~\cite{bib:coqDoc} terms.
We want to automatically identify \textit{goal clones}, \ie, \alphaEquivalent goals.
Such goals, and their corresponding sub-proofs, can be factored out as independent lemmas and reused, instead of proving them from scratch each time they are observed.
To solve this, we introduce \tool, a novel technique that leverages the formal definition of \alphaEquivalence of Gallina terms to systematically identify duplicated proof goals across large Rocq codebases.
This enables proof engineers to refactor their developments by extracting common subproofs as reusable lemmas, thereby reducing redundant proof effort and improving maintainability.
Although the problem statement is simple, and the notion of \alphaEquivalence of Gallina terms has a solid foundation in the well-established field of type theory, finding \alphaEquivalent Gallina terms in practice is not trivial.
This is mainly because the variables mean different things in different contexts and any comparison must first universally quantify the terms with the free variables defined in the context before performing \alphaEquivalence check (see~\cref{sec:approach} for more details on the proposed approach).

Our prototype Python implementation of \tool interacts with Rocq through Coq-LSP~\cite{bib:coqLSP} and its Python interface, CoqPyt~\cite{bib:carrott2024coqpyt}.
We evaluate \tool on 40 real-world Rocq projects from the CoqGym dataset~\cite{bib:yang2019coqgym}.
Our benchmark contains a diverse set of Rocq projects, representative of the real-world Rocq projects: the project are of different sizes, ranging from 53 to 29,260 lines of code, and they are from various
domains--from set theory to distributed system verification to geometry (see~\cref{sec:exp:benchmark} for more details on benchmark preparation process).
We ran \tool on our benchmark of Rocq projects, while measuring the number of goal clones that it detects as well as its runtime overhead.
We observed that in 20 out of 40 projects, \tool detected at least one goal clone.
On average \tool finds 27.73 instances of goal clones per project.
Manual review of the reported clones revealed three common patterns of goal cloning: (1) exact goal duplication with identical or near-identical proofs, \eg, structurally identical proofs with different variable names; (2) \alphaEquivalent goals with generalized proofs, where one proof subsumes the other; and (3) \alphaEquivalent goals with entirely different proofs.
Each of these instances signifies extra spent proof effort that could be saved and repurposed for other tasks by factoring out the duplicated goals as independent lemmas.
These findings suggest significant untapped potential for proof reuse in Rocq-based formal verification projects.

Another observation that we make is that \tool is a light-weight system: over 94.71\% of its running time is from using Coq-LSP through CoqPyt.
An average 45.31 seconds of run time, for a fresh run, and support for incremental analysis, which could significantly reduce the run time, make \tool a practical tool for daily proof engineering.
Mitigating the overhead of interfacing Coq-LSP is a matter of investing engineering effort so that CoqPyt does not reload common dependencies of the proof scripts within a workspace every time we load a script from that workspace.
Another possibility is to implement \tool directly as a Rocq plugin using Rocq plugin library and Coq-Elpi~\cite{bib:coqElpi}.
We save these as a future extension of this work.

Beyond detecting proof redundancy, our work contributes an open-source enhancement to CoqPyt, which enables analyzing real-world Rocq projects with it.
We will make both \tool and our enhancement publicly available to foster further research in automated proof engineering.
To summarize, this paper makes the following contributions.
\begin{itemize}
    \item \textbf{Technique:} a simple technique, named \tool, based on the theoretically elegant notion of \alphaEquivalence is proposed for detecting goal clones in Rocq scripts. This technique can be used alongside lemma extraction tool-set~\cite{bib:CompanyCoq2016,bib:roe2016coqpie} to rewrite clones as independent lemmas to increase reuse and mitigate waste of proof efforts. Additionally, the support for incremental analysis would allow using \tool in a background thread in a Rocq IDE to detect clones as the proof engineer writes the proofs.
    \item \textbf{Evaluation:} We evaluate \tool using a set of real-world Rocq projects. We report that real-world Rocq projects are likely to contain large numbers of clones with either (1) identical; (2) generalized; or (3) entirely different pairs of proofs. These clones indicate waste of manpower in proof engineering that could potentially be repurposed for use in other aspects of system specification and verification.
    \item \textbf{Open-Source Contribution:} While implementing \tool, which interacts with the Python interface for Coq-LSP, named CoqPyt, we realized that CoqPyt does not allow passing physical-to-logical path mapping, which is essential in working with real-world Rocq projects. We have added this feature to the framework and will make a pull request about it. We have also made \tool publicly available~\cite{bib:replica}. This source code contains a Python-based parser for Gallina that can be used in future research projects.
\end{itemize}

The remainder of this paper is organized as follows.
We provide background on Rocq and code clone detection in~\cref{sec:background}.
In~\cref{sec:approach} we present the \tool approach in detail, and in~\cref{sec:implementation} we discuss the implementation of our prototype.
We present our experimental evaluation in~\cref{sec:experiments}, followed by an analysis of threats to validity in~\cref{sec:threats}.
We review related work in~\cref{bib:related} and conclude the paper, while sketching future research directions, in~\cref{sec:confuture}.

\section{Background}\label{sec:background}
\begin{figure}[t!]
    \centering
    \includegraphics[scale=0.5]{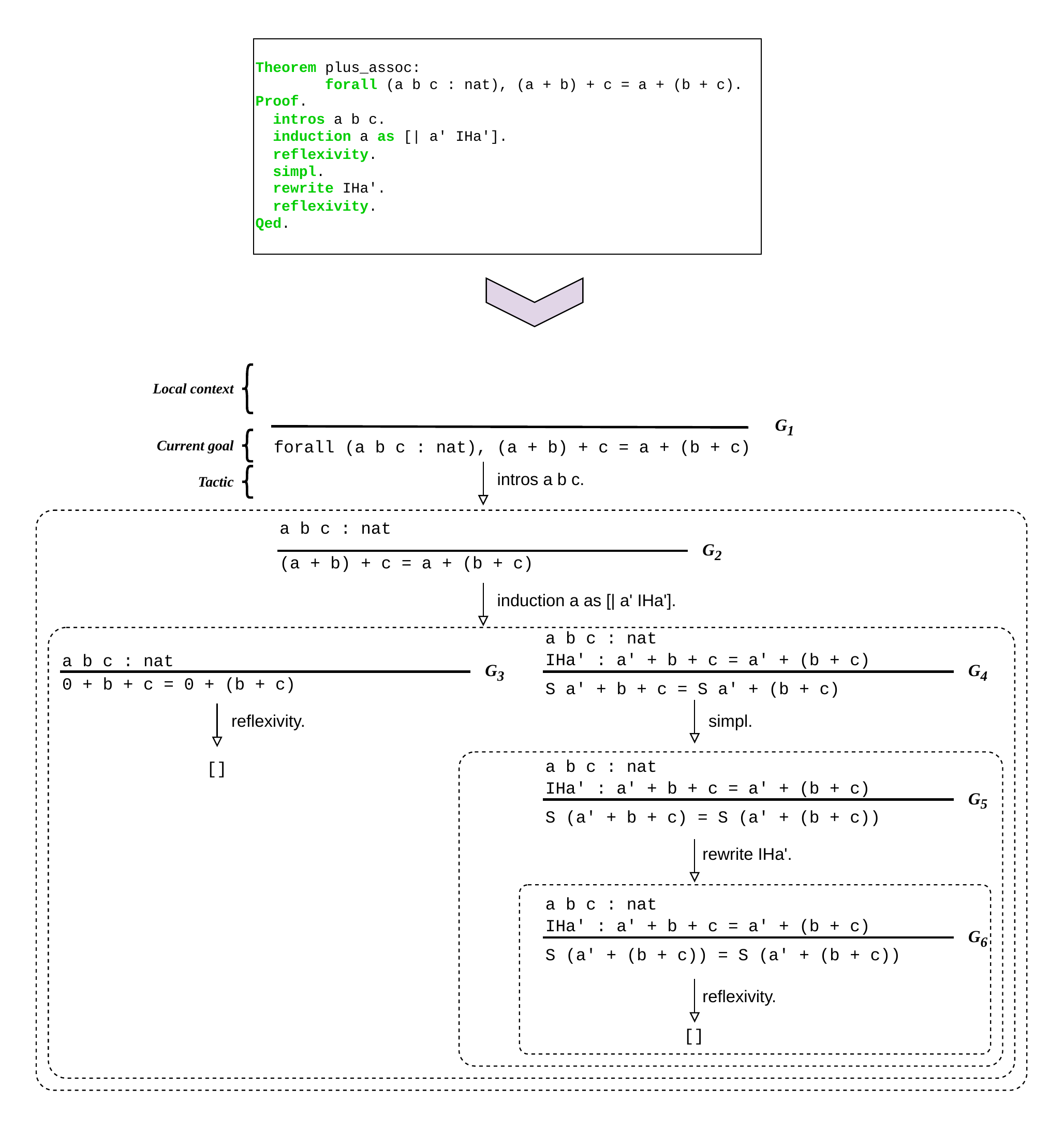}
    \caption{A simple Rocq theorem and its proof in Ltac. The proof process starts with the theorem statement as the current goal. Each valid Ltac command transforms the state of the proof assistant, prompting it to generate zero or more new goals. This amounts to creating a proof tree, where nodes corresponds to goals and the edges are labeled with the Ltac command that generated the child nodes. A depth-first search on each node of the tree to accumulate the tactics at the edges constitute the sub-proof for the goal at each node. This figure also illustrates how Algorithm~\ref{alg:proof_tree_algo} constructs proof trees by executing proofs line by line, while tracking the goals that are created and eliminated as a result of running each tactic}
    \label{fig:exampleProofTree}
\end{figure}
In this section, we briefly introduce Rocq, Gallina, Calculus of Inductive Constructions, and code clone detection.
We also provide references for further reading on these topics.

\subsection{Rocq, Gallina, and Calculus of Inductive Constructions}\label{sec:back:coq}
Rocq~\cite{bib:coqDoc}, formerly known as Coq, is an interactive theorem prover.
At its core, lies a specification language, named Gallina~\cite{bib:coqDoc}.
Gallina is an elegant functional programming language that was originally built upon the calculus of inductive constructions with definitions~\cite{bib:bertot2013interactive,bib:paulin1993inductive,bib:severi1994pure}.
This calculus still constitutes a substantial subset of Gallina today.
Proofs in Rocq are usually written in a tactic language, named Ltac~\cite{bib:delahaye2000ltac,bib:coqDoc}, which is a domain-specific language for building Gallina terms in a type-driven fashion that serve as proofs.
The type of these terms are also Gallina terms, that serve as theorem statements and goals.

Rocq is a state-based interactive proof assistant.
Theorem proving in Rocq starts with the theorem itself as the initial goal.
The proof engineer repeatedly applies Ltac tactics to decompose the goal into a list of subgoals.
The proof is complete when there are no sub-goals left.
At each step, Rocq displays the local definitions as available hypotheses that can be used in theorem proving.
A successful Rocq proof implicitly amounts to creating a proof tree whose root is the original theorem and whose nodes are the goals.
Figure~\ref{fig:exampleProofTree} shows a simple example.
All goals share the same global context, but have exclusive local contexts, \eg, hypotheses \textsf{a, b : nat} in Figure~\ref{fig:exampleProofTree}.
The edges of the proof tree are annotated with tactics.
The proof for a goal in the proof tree can be obtained by conducting a depth-first search~\cite{bib:algo2009clrs} on the node corresponding to the goal and accumulating the tactics that are visited along the way.

\subsection{Code Clone Detection}\label{sec:back:clone}
Code clone detection is an active area of research in software engineering aimed at identifying duplicated code snippets within software systems~\cite{micheline2024llm,bib:roy2007survey,bib:zakeri2023systematic,bib:white2016deep}.
Code duplication is widespread in software development.
It is sometimes introduced deliberately to accelerate prototyping, reuse existing functionality, or meet tight deadlines, and other times it arises inadvertently.
However, excessive cloning can degrade code quality, increase maintenance overhead, and propagate defects across multiple locations in the codebase~\cite{bib:elmar2009code,bib:kapser2008cloning}.

Code clones refer to code snippets that exhibit similarity based on a defined metric.
The literature categorizes code clones into four primary types~\cite{bib:roy2007survey,bib:zakeri2023systematic}: (1) exact clones; (2) renamed clones; (3) near-miss clones; and (4) semantic clones.

While the code clone detection technique introduced in this paper falls under the categories of type 1 and type 2 clones (our technique finds \alphaEquivalent goals in a project), the type of code that we study in this paper are fundamentally different from that of studied in the literature.
We are finding clones in functional programs, rather than imperative programs, so the existing techniques for code clone detection ~\cite{micheline2024llm,bib:zakeri2023systematic,bib:white2016deep} could not be used directly.

\section{The \tool Approach}\label{sec:approach}
At the highest level, \tool takes, as input, the base path for a Rocq project, Rocq options, and the minimum proof size threshold, and outputs a list of clones with proofs at least as large as the provided threshold.
The proof size is defined in terms of lines of Ltac code.
The operations of \tool can be summarized as (1) constructing proof trees (described in Algorithm~\ref{alg:proof_tree_algo}); (2) flattening, deduplicating, and generalization of the proof trees to construct the list of pairs of goals and their corresponding proofs (described in Algorithms~\ref{alg:flatten_algo} and~\ref{alg:generalize_algo}); (3) finding pairs of \alphaEquivalent generalized goals and reporting the output (described in Algorithm~\ref{alg:report_alpha_algo}).
We now describe each of these steps in more details.

\begin{figure}[t]
    \centering
    \resizebox{\textwidth}{!}{
    \begin{tabular}{|c|}
        \hline\\
    \[
        \begin{array}{c}
            \multicolumn{1}{c}{
                \begin{array}{rcl}
                    t & ::= & s \mid x \mid \cicApp{t_1}{t_2} \mid \cicProd{x}{t_1}{t_2} \\
                    & & \mid \cicFun{x}{t_1}{t_2} \mid \cicDef{x}{t_1}{t_2}{t_3} \\
                    & & \mid \cicFix{y}{x}{t_1}{t_2}{t_3} \\
                    & & \mid \cicMatch{t_1}{t_2}{~[\mathtt{|} x_1~\dots~x_n~\Rightarrow t_3]^*}
                \end{array}
            } \\
            \\
            \begin{array}{rrcl}
                \mbox{(terms)} & t, t_1, t_2, t_3 & \in & \mathbb{T} \\
                \mbox{(sorts)} & s & \in & \{\cicSet, \cicType\} \\
                \mbox{(variables)} & x, y, x_1, \dots, x_n & \in & \mathbb{V}
            \end{array}
        \end{array}
    \]
    \\\\\hline
    \end{tabular}}
    \caption{Abstract syntax of a core subset of Gallina terms.}
    \label{fig:syntax}
\end{figure}
The enabling tool of \tool are the notions of free variables and \alphaEquivalence.
To define these notions more precisely, we use a core subset of Gallina subsumed by the calculus of inductive constructions with definitions.
Figure~\ref{fig:syntax} shows this subset of Gallina.
Note that while inductive type definitions are part of the syntax of the calculus of inductive constructions with definitions, we have not included them in our core language, because this is how Rocq operates: after an inductive type definition is type checked, the type name, its constructors, as well as induction principles and recursor, are added as constants into the global type environment and type definition themselves do not appear in goals.
The following definitions define notions of free variables, substitution, and \alphaEquivalence.
Readers familiar with these standard notions may skip the following three definitions.

\begin{definition}[Free Variables]\label{def:fv}
The set of free variables of a Gallina term $t$, denoted $FV(t)$, is defined recursively as follows.
\begin{itemize}
    \item $FV(s)=\emptyset$, for all sorts $s$,
    \item $FV(x)=\{x\}$, for all $x\in\mathbb{V}$,
    \item $FV(\cicApp{t_1}{t_2})=FV(t_1)\cup FV(t_2)$ for all $t_1, t_2\in\mathbb{T}$,
    \item $FV(\cicProd{x}{t_1}{t_2})=FV(t_1)\cup (FV(t_2)-\{x\})$, for $x\in\mathbb{V}$ and $t_1, t_2\in\mathbb{T}$,
    \item $FV(\cicFun{x}{t_1}{t_2})=FV(t_1)\cup (FV(t_2)-\{x\})$, for $x\in\mathbb{V}$ and $t_1, t_2\in\mathbb{T}$,
    \item $FV(\cicDef{x}{t_1}{t_2}{t_3})=FV(t_1)\cup FV(t_2)\cup (FV(t_3)-\{x\})$, for $x\in\mathbb{V}$ and $t_1, t_2, t_3\in\mathbb{T}$,
    \item $FV(\cicFix{y}{x}{t_1}{t_2}{t_3})=FV(t_1)\cup FV(t_2)\cup (FV(t_3)-\{x, y\})$,
    \item $FV(\cicMatch{t_1}{t_2}{\mathtt{|} C_1\mathtt{|}\dots \mathtt{|} C_k})=FV(t_1)\cup FV(t_2)\cup FV_C(C_1)\cup \dots\cup FV_C(C_k)$, for $k\geq 0$,
\end{itemize}
where $s$ is a sort, $x, y\in\mathbb{V}$, and $t_1, t_2, t_3\in\mathbb{T}$. The helper function $FV_C$ is defined to be $FV_C(x_1~\dots~x_n~\Rightarrow t_3) = FV(t_3) - \{x_1,\dots,x_n\}$, wherein $n\geq 1$ and $x_1, \dots, x_n\in\mathbb{V}$.
\hfill\rule{1.2ex}{1.2ex}
\end{definition}

All the non-free variables occurring in a Gallina term are referred to as \textit{bound variables}.
Two Gallina terms are said to be \alphaEquivalent, if we can systematically rename bound variables in one term to make it syntactically equal to another.
To define this notion precisely, we first need to define variable replacement, which is used for variable renaming in the \alphaEquivalence check.

\begin{definition}[Substitution]
Given a variable $x$ and a term $u$, $\rep{t}{x}{u}$ denotes the substitution of $x$ with $u$ in $t$, and is defined as follows.
\begin{itemize}
    \item $\rep{s}{x}{u}=s$, for all sorts $s$,
    \item $\rep{x_0}{x}{u}=u$, if $x_0=x$,
    \item $\rep{x_0}{x}{u}=x_0$, for all $x_0\in\mathbb{V}$ such that $x_0\neq x$,
    \item $\rep{(\cicApp{t_1}{t_2})}{x}{u}=\cicApp{\rep{t_1}{x}{u}}{\rep{t_2}{x}{u}}$, for all $t_1, t_2\in\mathbb{T}$,
    \item $\rep{(\cicProd{x_0}{t_1}{t_2})}{x}{u}=\cicProd{x_0}{\rep{t_1}{x}{u}}{\rep{t_2}{x}{u}}$, for all $x_0\in\mathbb{V}$ and $t_1, t_2\in\mathbb{T}$ such that $x_0\neq x$ and $x_0\not\in FV(u)$,
    \item $\rep{(\cicProd{x_0}{t_1}{t_2})}{x}{u}=\cicProd{x_0}{\rep{t_1}{x}{u}}{t_2}$, if $x_0\not\in FV(u)$,
    \item $\rep{(\cicFun{x_0}{t_1}{t_2})}{x}{u}=\cicFun{x_0}{\rep{t_1}{x}{u}}{\rep{t_2}{x}{u}}$, for all $x_0\in\mathbb{V}$ and $t_1, t_2\in\mathbb{T}$ such that $x_0\neq x$ and $x_0\not\in FV(u)$,
    \item $\rep{(\cicFun{x_0}{t_1}{t_2})}{x}{u}=\cicFun{x_0}{\rep{t_1}{x}{u}}{t_2}$, for all $t_1, t_2\in\mathbb{T}$ such that $x_0\in FV(u)$,
    \item $\rep{(\cicDef{x_0}{t_1}{t_2}{t_3})}{x}{u}=\cicDef{x_0}{\rep{t_1}{x}{u}}{\rep{t_2}{x}{u}}{\rep{t_3}{x}{u}}$, for all $x_0\in\mathbb{V}$ and $t_1, t_2, t_3\in\mathbb{T}$ such that $x_0\neq x$ and $x_0\not\in FV(u)$,
    \item $\rep{(\cicDef{x_0}{t_1}{t_2}{t_3})}{x}{u}=\cicDef{x_0}{\rep{t_1}{x}{u}}{\rep{t_2}{x}{u}}{t_3}$, for all $t_1, t_2, t_3\in\mathbb{T}$ such that $x_0\in FV(u)$,
    \item $\rep{(\cicFix{y}{x_0}{t_1}{t_2}{t_3})}{x}{u}=\cicFix{y}{x_0}{\rep{t_1}{x}{u}}{\rep{t_2}{x}{u}}{\rep{t_3}{x}{u}}$, for all $y, x_0\in\mathbb{V}$ and $t_1, t_2, t_3\in\mathbb{T}$ such that $y\neq x$, $x_0\neq x$, $x_0\not\in FV(u)$, and $y\not\in FV(u)$,
    \item $\rep{(\cicFix{y}{x_0}{t_1}{t_2}{t_3})}{x}{u}=\cicFix{y}{x_0}{\rep{t_1}{x}{u}}{\rep{t_2}{x}{u}}{t_3}$, for all $t_1, t_2, t_3\in\mathbb{T}$ such that $x_0\in FV(u)$ or $y\in FV(u)$,
    \item $\rep{(\cicMatch{t_1}{t_2}{\mathtt{|} C_1\mathtt{|}\dots \mathtt{|} C_k})}{x}{u}= \cicMatch{\rep{t_1}{x}{u}}{\rep{t_2}{x}{u}}{\mathtt{|} \rep{C_1}{x}{u}\mathtt{|}\\\dots \mathtt{|} \rep{C_k}{x}{u}}$, where $\rep{C_i}{x}{u}$ is defined as $\rep{(x_1~\dots~x_n~\Rightarrow t_3)}{x}{u}=(x_1~\dots~x_n~\Rightarrow \rep{t_3}{x}{u})$, if $x\not\in\{x_1,...,x_n\}$ and $\{x_1,...,x_n\}\cap FV(u)=\emptyset$, and $\rep{(x_1~\dots~x_n~\Rightarrow t_3)}{x}{u}=x_1~\dots~x_n~\Rightarrow t_3$ if $\{x_1,...,x_n\}\cap FV(u)\neq\emptyset$.
\end{itemize} 
\hfill\rule{1.2ex}{1.2ex}
\end{definition}

We now define the notion of \alphaEquivalence as follows.
\begin{algorithm}[t!]
    \caption{Constructing proof trees from Rocq proofs} \label{alg:proof_tree_algo}
    \begin{scriptsize}
        \begin{algorithmic}[1]
            \Require List of Rocq source files, $\text{coq\_files}$, and \texttt{coqc} flags, stored in variable `$\text{flags}$'
            \Ensure Returning a mapping from $\langle\text{File}, \text{Theorem}\rangle$ to $\text{ProofTree}$
            
            \State $\text{tree\_map} \gets \emptyset$
            \ForAll{$\text{file} \in \text{coq\_files}$} \Comment{Construct a proof tree for each theorem in each file}
                \State $\text{pf} \gets \text{connect\_to\_coq\_lsp}(\text{file}, \text{flags})$ \Comment{Connect to Coq-LSP: an iterable proof file object is returned}
                \ForAll{$\text{step} \in \text{pf}$} \Comment{Sweep the current file step-by-step}
                    \State $\text{step.execute()}$
                    \If{$\text{step.in\_proof()}$} \Comment{Have stepped into the proof of a theorem?}
                        \State $\text{thm} \gets \text{step.get\_theorem()}$ 
                        \State $\text{tree} \gets \text{ProofTree()}$ \Comment{Initialize a new proof tree}
                        \State $\text{tree\_map}[\langle\text{file}, \text{thm}\rangle] \gets \text{tree}$ \Comment{Store the proof tree}
                        \State $\text{prev} \gets \emptyset$ \Comment{Track goals at the last step}
                        
                        \While{\textbf{not} $\text{step.can\_close()}$} \Comment{Iterate through the proof steps until no more goals remain}
                            \State $\text{step.execute()}$
                            
                            \State $\text{curr} \gets \emptyset$ \Comment{Store current proof state}
                            \ForAll{$\text{g} \in \text{pf.goals()}$}
                                \State $\text{ctx} \gets \text{g.get\_hypotheses()}$ \Comment{Extract local and in-file context (hypotheses)}
                                \State $\text{goal} \gets \text{g.get\_type()}$ \Comment{Extract the goal type (\textit{i.e.}, goal statement)}
                                \State $\text{curr} \gets \text{curr} \cup \{\langle\text{ctx}, \text{goal}\rangle\}$ \Comment{Store current goals}
                            \EndFor
                            
                            \If{$\text{prev} \neq \emptyset$}
                                \State $\text{gone} \gets \{t \in \text{prev} \mid t \notin \text{curr}\}$ \Comment{Goals that disappeared}
                                \State $\text{new} \gets \{t \in \text{curr} \mid t \notin \text{prev}\}$ \Comment{Newly introduced goals}
                                
                                \ForAll{$t \in \text{gone}$}
                                    \State $\text{ctx} \gets t[0]$
                                    \State $\text{goal} \gets t[1]$
                                    \State $\text{children} \gets \{c[1] \mid c \in \text{new}\}$ \Comment{Children are new goals}
                                    \State $\text{tac} \gets \text{step.get\_tactic()}$ \Comment{Get the applied tactic}
                                    \State $\text{tree.AddNode}(\text{goal}, \text{ctx}, \text{children}, \text{tac})$\\\Comment{Add node to proof tree and connect it to its children}
                                \EndFor
                            \EndIf
                            
                            \State $\text{prev} \gets \text{curr}$ \Comment{Update tracking state}
                        \EndWhile
                    \EndIf
                \EndFor
                \State \text{close}(\text{pf}) \Comment{Close the proof file and the Coq-LSP connection}
            \EndFor
            \State \Return $\text{tree\_map}$ \Comment{Return the constructed proof tree map}
        \end{algorithmic}
    \end{scriptsize}
\end{algorithm}

\begin{definition}[\alphaEquivalence]
The following axiom and rules of inference prove judgments of the form $\aEq{t_1}{t_2}$, where $t_1$ and $t_2$ are Gallina terms.
\begin{itemize}
    \item \textbf{(Renaming-Forall)}: $\aEq{(\cicProd{x}{t_1}{t_2})}{(\cicProd{y}{t'_1}{t'_2})}$, if $\aEq{t_1}{t'_1}$, $y\not\in FV(t_2)$, $y$ is not bound in $t_2$, and $\aEq{\rep{t_2}{x}{y}}{t'_2}$,
    \item \textbf{(Renaming-Fun)}: $\aEq{(\cicFun{x}{t_1}{t_2})}{(\cicFun{y}{t'_1}{t'_2})}$, if $\aEq{t_1}{t'_1}$, $y\not\in FV(t_2)$, $y$ is not bound in $t_2$, and $\aEq{\rep{t_2}{x}{y}}{t'_2}$,
    \item \textbf{(Renaming-Let)}: $\aEq{(\cicDef{x}{t_1}{t_2}{t_3})}{(\cicDef{y}{t'_1}{t'_2}{t'_3})}$, if $\aEq{t_1}{t'_1}$, $\aEq{t_2}{t'_2}$, $y\not\in FV(t_3)$, $y$ is not bound in $t_3$, and $\aEq{\rep{t_3}{x}{y}}{t'_3}$,
    \item \textbf{(Renaming-Fix)}: $\aEq{(\cicFix{y}{x}{t_1}{t_2}{t_3})}\cicFix{y'}{x'}{t'_1}{t'_2}{t'_3}$, if $t_1=_\alpha t'_1$, $t_2=_\alpha t'_2$, $y'\not\in FV(t_3)$, $y'$ is not bound in $t_3$, $x'\not\in FV(t_3)$, $x'$ is not bound in $t_3$, and $t_3^{y\rightarrow y'~x\rightarrow x'}=_\alpha t'_3$,
    \item \textbf{(Renaming-Match)}: $(\mathtt{match}~t_1~\mathtt{return}~t_2~\mathtt{with}~|~C_1~|~\dots~|~C_k~\mathtt{end})\\=_\alpha (\mathtt{match}~t'_1~\mathtt{return}~t'_2~\mathtt{with}~|~C'_1~|~\dots~|~C'_k~\mathtt{end})$, if $t_1=_\alpha t'_1$, $t_2=_\alpha t'_2$, and for all $1\leq i\leq k$, we have $C_i\equiv_\alpha C'_i$, where $\equiv_\alpha$ is defined as: $(x_1~\dots~x_n~\Rightarrow t_3)\equiv_\alpha (x'_1~\dots~x'_n~\Rightarrow t'_3)$ if $x'_1~\dots~x'_n\not\in FV(t_3)$, variables $x'_1~\dots~x'_n$ are not bound in $t_3$, and $t_3^{x_1\rightarrow x'_1~\dots~x_n\rightarrow x'_n}=_\alpha t'_3$.
    \item \textbf{(Compatibility-1)}: If $t_1=_\alpha t_2$, then $t_1~t=_\alpha t_2~t$, $t~t_1=_\alpha t~t_2$, $(\mathtt{forall}~(x : t_1), t)=_\alpha (\mathtt{forall}~(x : t_2), t)$, and $(\mathtt{fun}~(x : t_1)\Rightarrow t)=_\alpha (\mathtt{fun}~(x : t_2)\Rightarrow t)$, for all term $t$,
    \item \textbf{(Compatibility-2)}: If $t_1=_\alpha t_2$, then $(\mathtt{match}~t_1~\mathtt{return}~t~\mathtt{with}~|~C_1~|~\dots~|~C_k~\mathtt{end})=_\alpha (\mathtt{match}~t_2~\mathtt{return}~t~\mathtt{with}~|~C_1~|~\dots~|~C_k~\mathtt{end})$ and $(\mathtt{match}~t~\mathtt{return}~t_1~\mathtt{with}~|~C_1~|~\dots\\|~C_k~\mathtt{end})=_\alpha (\mathtt{match}~t~\mathtt{return}~t_2~\mathtt{with}~|~C_1~|~\dots~|~C_k~\mathtt{end})$, for all term $t$ and match clauses $C_1$, ..., $C_k$.
    \item \textbf{(Compatibility-3)}: If $t_1=_\alpha t_2$, then $(\mathtt{let}~x : t_1 := t~\mathtt{in}~u)=_\alpha (\mathtt{let}~x : t_2 := t~\mathtt{in}~u)$, $(\mathtt{let}~x : t := t_1~\mathtt{in}~u)=_\alpha (\mathtt{let}~x : t := t_2~\mathtt{in}~u)$, $(\mathtt{fix}~y (x : t_1): t\Rightarrow u)=_\alpha (\mathtt{fix}~y (x : t_2): t\Rightarrow u)$, and $(\mathtt{fix}~y (x : t): t_1\Rightarrow u)=_\alpha (\mathtt{fix}~y (x : t): t_2\Rightarrow u)$, for all terms $t$ and $u$,
    \item \textbf{(Compatibility-4)}: If $t_1=_\alpha t_2$, then $(\mathtt{forall}~(x : t), t_1)=_\alpha (\mathtt{forall}~(z : t), t_2)$, $(\mathtt{fun}~(x : t)\Rightarrow t_1)=_\alpha (\mathtt{fun}~(z : t)\Rightarrow t_2)$, for all term $t$ and variable $z$,
    \item \textbf{(Compatibility-5)}: If $t_1=_\alpha t_2$, then $(\mathtt{let}~x : t := u~\mathtt{in}~t_1)=_\alpha (\mathtt{let}~z : t := u~\mathtt{in}~t_2)$, $(\mathtt{fix}~y (x : t): u\Rightarrow t_1)=_\alpha (\mathtt{fix}~z (x : t): u\Rightarrow t_2)$, and $(\mathtt{fix}~y (x : t): u\Rightarrow t_1)=_\alpha (\mathtt{fix}~y (z : t): u\Rightarrow t_2)$, and $(\mathtt{match}~t~\mathtt{return}~u~\mathtt{with}~|~C_1~|~\dots|(x_1~\dots~x~\dots~x_n)\Rightarrow t_1~\dots~|~C_k~\mathtt{end})=_\alpha (\mathtt{match}~t~\mathtt{return}~u~\mathtt{with}~|~C_1~|~\dots|(x_1~\dots~z~\dots~x_n)\Rightarrow t_2~\dots\\|~C_k~\mathtt{end})$, for all terms $t$ and $u$, and variable $z$,
    \item \textbf{(Reflexivity)}: $t=_\alpha t$, for all term $t$,
    \item \textbf{(Symmetry)}: If $t_1=_\alpha t_2$ then $t_2=_\alpha t_1$,
    \item \textbf{(Transitivity)}: If $t_1=_\alpha t_2$ and $t_2=_\alpha t_3$ then $t_1=_\alpha t_3$.
\end{itemize}
Renaming rules are the foundations of $\alpha$-equivalence. Compatibility rules have the effect of extending $\alpha$-equivalence from subterms to bigger terms. Reflexivity, symmetry, and transitivity make $\alpha$-equivalence an equivalence relation.
If we can derive $\aEq{t_1}{t_2}$, we say that the terms $t_1$ and $t_2$ are \alphaEquivalent.
\hfill\rule{1.2ex}{1.2ex}
\end{definition}

Armed with these definitions, we are now in a position to define our algorithms making up \tool.
The first step is to extract goals and construct proof trees, which is described in Algorithm~\ref{alg:proof_tree_algo}.
This step is performed to find the proof corresponding to each goal in a Rocq project.
\tool invokes this algorithm by first finding all the \texttt{.v} files in the target Rocq project directory.
This list, together with \texttt{coqc} flags are passed to the algorithm.
In summary, this algorithm runs each Rocq script in the given project step by step to construct proof trees using a tree data structure.
Proof tree construction is done as follows.
After executing a tactic, the current goal disappears, and a set of new goals might be created, these goals are regarded as the children of the current goal in the proof tree.
We can identify the edges of the tree by tracking how goals are created during the proof execution.
For example, if the list of goals changes from $[A, B]$ to $[C, D, B]$, we know that node $A$ has two children $C$ and $D$.
Algorithm~\ref{alg:proof_tree_algo} creates nodes that point to the children node with the help of a dictionary-based data structure that maps current goal to the child goals and other information about the current goals such as the tactic that transforms it to the child goals and the context right before the execution of the tactic.
Figure~\ref{fig:exampleProofTree} provides a simple example of a Rocq proof and its corresponding proof tree.

More precisely, Algorithm~\ref{alg:proof_tree_algo} constructs a mapping of Rocq scripts to their corresponding proof trees, capturing the hierarchical structure of goals and tactics used to prove them.
It iterates over each Rocq source file (line 2), establishing a connection to Coq-LSP to analyze proof steps (lines 3 and 4).
When a proof of a theorem is encountered (line 6), the algorithm initializes a proof tree data structure and adds it to the resulting map, mapping file name and the the current theorem name pairs to the newly created proof tree (lines 7-9).
This proof tree object, gets populated as follows.
The algorithm iterates through proof steps until the proof is complete (line 11).
At each step, it extracts the current goals and the local, and in-file, contexts (lines 13–18), detects goals that have disappeared or emerged (lines 20–21), and records the proof structure by adding nodes to the proof tree (lines 22–29).
Finally, after processing all steps, the proof file is closed (line 35), and the mapping of file names and theorem names to proof trees is returned (line 31).
\begin{algorithm}[t!]
    \caption{Flatten and remove redundant goals} \label{alg:flatten_algo}
    \begin{scriptsize}
        \begin{algorithmic}[1]
            \Require pt\_map, a mapping from $\langle\text{File}, \text{Theorem}\rangle$ to $\text{ProofTree}$, \ie, Algorithm~\ref{alg:proof_tree_algo}'s output
            \Ensure List of goals and their generalizations, where redundant goals have been removed
            \State $\text{g\_list} \gets []$ \Comment{Initialize list for storing goals}
            
            \ForAll{$(\text{f}, \text{t}, \text{pt}) \in \text{pt\_map}$} \Comment{Iterate over proof tree map}
                \ForAll{$\text{g} \in \text{pt.GetNodes()}$} \Comment{Extract goals from proof tree}
                    \State $\text{ctx} \gets \text{g.GetContext()}$
                    \State $\text{g\_gen} \gets \text{Generalize(ctx, g)}$
                    \State $\text{p} \gets \text{g.GetProof()}$
                    \State $\text{g\_list.append}(\langle \text{g}, \text{g\_gen}, \text{p}, \text{f}, \text{t} \rangle)$
                \EndFor
            \EndFor
            
            \State $\text{red\_g} \gets \{\}$ \Comment{Initialize set for redundant goals}
            
            \For{$\text{i} = 0$ to $|\text{g\_list}| - 1$} \Comment{Iterate through goals list}
                \For{$\text{j} = 0$ to $|\text{g\_list}| - 1$} \Comment{Compare each goal with others}
                    \If{$\text{i}\neq\text{j}$ and $\text{prodBody(g\_list[i][1], g\_list[j][1])}$} \Comment{Check if one goal is part of another}
                        \State $\text{red\_g.add(g\_list[j][1])}$
                    \EndIf
                \EndFor
            \EndFor
            
            \State $\text{g\_list} \gets [\text{g} \in \text{g\_list} \mid \text{g}[1] \notin \text{red\_g}]$ \Comment{Remove redundant goals}
            \State \Return $\text{g\_list}$
        \end{algorithmic}
    \end{scriptsize}
\end{algorithm}

The next step in \tool's pipeline is to flatten the proof trees by turning the proof tree map generated by Algorithm~\ref{alg:proof_tree_algo} into a list of tuples of a goal, its generalization (\ie, universal quantification of its free variables), its proof, its containing file name, and its containing theorem name.
At this step, \tool also deduplicates the goals by removing the tuples whose generalization is a body of another tuple's generalization.
Algorithm~\ref{alg:flatten_algo} formalizes this process.
This algorithm processes a proof tree map by flattening its structure and eliminating redundant goals.
First, it initializes an empty list g\_list (line 1) to store extracted goals and their associated information.
Then, it iterates through the proof tree map (pt\_map) and extracts goals from each proof tree (lines 2–3).
The method GetNodes of proof tree, returns the set of nodes, \ie, the goals, in the proof tree.
For each goal, it retrieves its context, generalizes it, and pairs it with its proof, file, and theorem name before appending this information to g\_list (lines 4–7).
The method GetProof, receives a node, \ie, a goal, as input, conducts a depth-first search rooted at that node, while accumulating the tactics along each edge that it visits.
The list of tactics ultimately returned by the method constitutes the proof for that goal.
Another function used by the algorithm is the Generalize function, which is calculated according to Algorithm~\ref{alg:generalize_algo} and will be described shortly.
Next, a set red\_g is initialized (line 10) to keep track of redundant goals.
The algorithm then performs a pairwise comparison of the goals (lines 11–13), marking a goal as redundant if its generalized form appears as a subterm within another goal using the function prodBody (line 13).
The definition below, provides a more formal description of the prodBody function.
\begin{algorithm}[t!]
    \caption{Generalization of a Goal} \label{alg:generalize_algo}
    \begin{scriptsize}
        \begin{algorithmic}[1]
            \Require $\text{ctx}$: A map from variables to terms
            \Require $\text{goal}$: The goal to be generalized
            \Ensure A generalized goal with universally quantified variables
            \State $\text{fvs} \gets \{ v \in FV(\text{goal}) \mid v \in \text{ctx} \}$ \Comment{Find free variables bound in the context}
            \State $\text{rev\_top\_order} \gets \text{reverse}(\text{topo\_sort}(\text{ctx}, \text{fvs}))$ \Comment{Sort dependencies in reverse topological order}
            \For{each $v \in \text{rev\_top\_order}$}
                \State $\text{goal} \gets \text{``\texttt{forall} (''} \, v \text{`` : ''} \, \text{ctx}[v] \text{``, ''} \, \text{goal} \text{``)''}$ 
                \Comment{Prepend universal quantifiers}
            \EndFor
            \State \Return $\text{goal}$
        \end{algorithmic}
    \end{scriptsize}
\end{algorithm}

\begin{definition}[prodBody Function]
    Given Gallina terms $t$ and $t'$, $\text{prodBody}(t,t')$ returns \textit{true} if $t$ is a product type, aka, universally quantified term, with body $t''$, \ie, $t=\texttt{forall}~(x:t_0),~t''$ for some term $t_0$, such that either $t'=t''$ or $\text{prodBody}(t'',t')$ is \textit{true}.
\end{definition}

Finally, the redundant goals are removed from g\_list (line 18), ensuring that only distinct, non-nested goals remain.

The Generalize function transforms a given \text{goal} into a universally quantified form by identifying and universally quantifying all free variables in the goal that are present in the context (\text{ctx}). 
Algorithm~\ref{alg:generalize_algo} defines this function.
This function first collects the relevant free variables (\text{fvs}) and determines their dependency order using \text{topo\_sort} (Line 2), ensuring that variables are introduced in a correct dependency-aware sequence. 
The topological sorting in reverse order ensures that if variable v1 depends on v2, then v2 is quantified before v1, preventing type dependency issues.
The function then iterates through the reverse of this topological order (Line 2) and prepends universal quantifiers (\texttt{forall}) to the goal in the correct order, making sure all dependent variables appear before their use. 
This process ensures that dependently typed terms remain well-formed while abstracting away unnecessary local dependencies.

The function \text{topo\_sort} constructs a topological ordering~\cite{bib:algo2009clrs} of variables in \text{ctx} that are also present in \text{fvs}.
It first builds a directed acyclic graph wherein nodes are variables and edges represent dependency relationships between the variables.
More precisely, assume that we have $v_1: t_1$ and $v_2: t_2$ in local context, where $t_1$ is a Gallina term such that $v_2\in FV(t_1)$.
Intuitively, this means that $v_1$ needs $v_2$ in order for it to be defined, because its type uses $v_2$, so we create a directed edge from the node corresponding to $v_1$ to that of $v_2$ to denote this dependency.
The function then conducts a topological sorting on this graph by conducting a depth-first search on each node.
The output of \text{topo\_sort} function is a list of variables that if reversed will present the variables in a valid dependency order.
So, if $v_1$ depends on $v_2$ in the graph, we will universally quantify $v_2$ before $v_1$, thereby creating a well-formed dependently typed term.

The last step in \tool's pipeline involves finding pairs of generalized goals that are \alphaEquivalent.
This process is formalized in Algorithm~\ref{alg:report_alpha_algo}.
The algorithm simply picks pairs of tuples from the output of Algorithm~\ref{alg:flatten_algo} and checks if the compartments corresponding to the generalized goals are \alphaEquivalent.
If so, the tuples, that contain various information, such as the containing theorem name, file name, and proofs, are concatenated to form the output of \tool.

%
\begin{algorithm}[t!]
    \caption{Reporting clones} \label{alg:report_alpha_algo}
    \begin{scriptsize}
        \begin{algorithmic}[1]
            \Require $\text{goals\_list}$: List of deduplicated and generalized goals, \ie, Algorithm~\ref{alg:flatten_algo}'s output
            \Ensure List of alpha equivalent goals 
            \State $\text{output} \gets \emptyset$
            \For{$i = 0~\text{to}~|\text{goals\_list}| - 1$}
                \For{$j = i + 1~\text{to}~|\text{goals\_list}| - 1$}
                    \If{$\text{goals\_list}[i][1]~\aEq~\text{goals\_list}[j][1]$}
                        \State $\text{output.append}(\text{goals\_list}[i] \cdot \text{goals\_list}[j])$ 
                        \State \Comment{Concatenate \text{goals\_list}[i] and \text{goals\_list}[j] to generate output pairs}
                    \EndIf
                \EndFor
            \EndFor
            \State \Return $\text{output}$
        \end{algorithmic}
    \end{scriptsize}
\end{algorithm}

\section{Implementation}\label{sec:implementation}
We have implemented \tool in 4,969 lines of Python code.
As we have given high-level explanation in~\cref{sec:approach}, \tool interacts with Coq-LSP to open the Rocq files in a project and execute them step by step, extracting goals, as well as local context entries.
In this implementation of \tool, for the sake of efficiency, we only consider local contexts, which could introduce unsoundness and false positives.
We have implemented a Gallina parser to parse the goals and find \alphaEquivalent goals, as outlined in~\cref{sec:approach}.
Currently, \tool offers a command-line interface though which it receives target project's base directory, Rocq command-line options, \ie, \texttt{-R}/\texttt{-Q} for mapping physical paths to logical paths, and the minimum size of the proofs.
\tool interfaces Coq-LSP through a recent Python framework, named CoqPyt~\cite{bib:carrott2024coqpyt}.
We observed that CoqPy does not allow passing \texttt{-R}/\texttt{-Q} options to Coq-LSP, which is essential for running real-world projects that we have studied in this paper.
Therefore, we extended the API provided by the framework so that they receive \texttt{-R}/\texttt{-Q} options and pass them to Coq-LSP.
We shall publish this open-source contribution.

Minimum size of proofs parameter of \tool dictates the tool to ignore clones with proofs under a certain number of lines of Ltac code.
This parameter defaults to 5, \ie, only clones with proofs equal to 5 lines of Ltac commands or more will be reported.
In~\cref{sec:experiments}, we study \tool's performance with this default value.
Studying \tool with different parameter sizes is the subject of a future extension of this work.
We would also like to emphasize that \tool could be implemented using Coq-Elpi~\cite{bib:coqElpi}.
Our Python implementation of Gallina parser calls for disabling the notations that might have been used in the proof scripts.
Specifically, in all our experiments, we invoked \tool by first prepending each proof script with the directive \texttt{Set Printing All} to disable the notations.
This is essential for our parser to work.
Despite this limitation, we opted for using Python and directly parsing Gallina expressions, as we believe that this approach would make this infrastructure available for a wider research community and foster research in automated theorem proving.

\section{Experiments}\label{sec:experiments}
In this section, we report our experimental results for running \tool on a number of real-world Rocq projects.
All our experiments are conducted on a Dell workstation with a 48-core Intel Xeon Gold CPU @ 2.3 GHz and 512 GB of RAM, running Ubuntu 22.04.5 LTS.

\subsection{Benchmark Projects}\label{sec:exp:benchmark}
Our benchmark suite consists of real-world Rocq projects from the CoqGym dataset~\cite{bib:yang2019coqgym}, which is a widely used dataset in the literature~\cite{bib:yang2019coqgym,bib:stern2023passport,bib:stern2020proverbot,bib:first2023baldur,bib:first2020tactok,bib:first2022diva}.
CoqGym comprises 135 Rocq projects obtained from GitHub.
We observed that the projects were compatible with 5 different versions of Rocq.
Older projects were compatible with Coq version 8.5, 8.9.0, or 8.12.0.
46 projects were compatible with version 8.16.0, and 36 projects with version 8.20.0, \ie, the latest version of Rocq as of conducting our experiments.
Coq 8.16.0 is the first version of Rocq that includes Coq-LSP, so we had to use version 8.16.0 or newer.
We chose Coq 8.16.0, because it was compatible with most number of projects.
Unfortunately, the older version of Coq-LSP, shipped with Coq 8.16.0, occasionally crashed with a ``segmentation fault'' error, when we used it through CoqPyt, so we ended up with 40 working Rocq projects.
\begin{table}[t!]
  \centering
  \caption{Benchmark of Rocq projects and the result of applying \tool on each project}\label{tab:mainTable}
    \resizebox{\textwidth}{!}{
        \begin{tabular}{|l|r|r|r|r|r|r|}
        \hline
        \multicolumn{1}{|c|}{\multirow{2}{*}{\textbf{Project Name}}} & \multicolumn{2}{c|}{\textbf{Project Size}} & \multicolumn{1}{r|}{\multirow{2}{*}{\textbf{\# Clones}}} & \multicolumn{3}{c|}{\textbf{Time (s)}} \\
    \cline{2-3}\cline{5-7}          & \multicolumn{1}{c|}{\textbf{\# Files}} & \multicolumn{1}{c|}{\textbf{LoC}} &       & \multicolumn{1}{c|}{\textbf{Build}} & \multicolumn{1}{c|}{\textbf{\tool Total}} & \multicolumn{1}{c|}{\textbf{Analysis}} \\
        \hline
        \hline
        \textsf{AlmostFull} & 11    & 3,625 & 102   & 8.563 & 1962.643 & 27.661 \\
        \hline
        \textsf{Angles} & 6     & 3,633 & 25    & 4.356 & 79.379 & 50.821 \\
        \hline
        \textsf{Autosubst} & 8     & 1,553 & 0     & 3.484 & 104.820 & 0.511 \\
        \hline
        \textsf{bbv} & 16    & 9,027 & 6     & 13.460 & 3137.301 & 93.807 \\
        \hline
        \textsf{CDF} & 13    & 8,558 & 33    & 14.127 & 3502.113 & 21.409 \\
        \hline
        \textsf{Chapar} & 10    & 20,371 & 447   & 303.367 & 2896.301 & 223.292 \\
        \hline
        \textsf{Checker} & 2     & 141   & 0     & 0.950 & 11.309 & 0.016 \\
        \hline
        \textsf{ConstructiveGeometry} & 7     & 1,453 & 0     & 2.123 & 21.533 & 0.204 \\
        \hline
        \textsf{Coqoban} & 2     & 6,398 & 0     & 2.286 & 16.486 & 0.947 \\
        \hline
        \textsf{DepMap} & 6     & 1,259 & 0     & 6.701 & 1538.239 & 5.151 \\
        \hline
        \textsf{DomainTheory} & 6     & 1,223 & 1     & 3.472 & 96.321 & 5.981 \\
        \hline
        \textsf{ExtLib} & 118   & 10,304 & 3     & 29.696 & 2860.590 & 9.163 \\
        \hline
        \textsf{FreeGroups} & 1     & 654   & 0     & 1.166 & 71.283 & 0.785 \\
        \hline
        \textsf{FunctionsInZFC} & 1     & 6,294 & 0     & 1.414 & 18.383 & 0.001 \\
        \hline
        \textsf{Game} & 15    & 1,345 & 1     & 2.950 & 101.722 & 1.158 \\
        \hline
        \textsf{Groups} & 1     & 306   & 0     & 0.850 & 8.788 & 0.089 \\
        \hline
        \textsf{GroupTheory} & 12    & 1,661 & 2     & 3.793 & 843.374 & 1.891 \\
        \hline
        \textsf{Hedges} & 1     & 3,265 & 8     & 2.775 & 164.673 & 32.239 \\
        \hline
        \textsf{HighSchoolGeometry} & 72    & 29,260 & 218   & 185.082 & 29773.385 & 1062.286 \\
        \hline
        \textsf{Huffman} & 22    & 6,791 & 9     & 15.160 & 3112.586 & 29.449 \\
        \hline
        \textsf{IZF} & 9     & 2,419 & 3     & 2.207 & 23.045 & 1.169 \\
        \hline
        \textsf{JML} & 30    & 20,009 & 27    & 23.594 & 5292.184 & 48.636 \\
        \hline
        \textsf{Lambda} & 12    & 2,122 & 0     & 5.559 & 1374.476 & 1.899 \\
        \hline
        \textsf{LibHyps} & 8     & 2,957 & 0     & 2.013 & 1111.587 & 0.000 \\
        \hline
        \textsf{mathcomp.bigenough} & 1     & 122   & 0     & 1.272 & 185.391 & 0.000 \\
        \hline
        \textsf{mathcomp.finmap} & 2     & 5,029 & 0     & 19.181 & 1204.847 & 3.433 \\
        \hline
        \textsf{mathcomp.zify} & 4     & 1,583 & 0     & 23.888 & 3616.383 & 0.025 \\
        \hline
        \textsf{MiniML} & 1     & 1,425 & 3     & 1.958 & 24.486 & 6.717 \\
        \hline
        \textsf{NotationGallery} & 1     & 53    & 0     & 0.120 & 93.005 & 0.000 \\
        \hline
        \textsf{OtwayRees} & 21    & 1,401 & 16    & 5.850 & 119.969 & 8.015 \\
        \hline
        \textsf{parseque} & 14    & 833   & 0     & 4.811 & 749.475 & 0.000 \\
        \hline
        \textsf{Ramsey} & 1     & 172   & 0     & 0.603 & 50.528 & 0.000 \\
        \hline
        \textsf{RecordUpdate} & 3     & 131   & 0     & 0.947 & 11.400 & 0.000 \\
        \hline
        \textsf{RSA} & 5     & 1,654 & 0     & 3.991 & 1010.601 & 13.068 \\
        \hline
        \textsf{Schroeder} & 5     & 640   & 0     & 1.503 & 76.689 & 0.573 \\
        \hline
        \textsf{Stalmarck} & 38    & 14,108 & 78    & 30.861 & 7819.962 & 134.503 \\
        \hline
        \textsf{Subst} & 20    & 7,506 & 82    & 11.001 & 132.391 & 9.668 \\
        \hline
        \textsf{WeakUpTo} & 10    & 2,334 & 7     & 3.433 & 26.097 & 2.284 \\
        \hline
        \textsf{ZFC} & 11    & 4,336 & 38    & 4.538 & 47.417 & 15.437 \\
        \hline
        \textsf{ZF} & 15    & 7,977 & 0     & 5.015 & 42.641 & 0.000 \\
        \hline
        \end{tabular}
    }
\end{table}

Columns ``Project Name'' and ``Project Size'' of Table~\ref{tab:mainTable} list all the projects in our benchmark and their sizes.
We have reported the project sizes both in terms of number of \texttt{.v} files in the project (column ``\#Files'') and total lines of code (column ``LoC'').
Our benchmark contains a diverse set of Rocq projects, representative of the real-world Rocq projects.
The projects are of varying sizes from 53 to 29,260 lines of code, and they are from various domains: from set theory to distributed system verification to geometry.

\subsection{Results}\label{sec:exp:results}
We ran \tool on each of our benchmark projects.
The last four columns of Table~\ref{tab:mainTable} report our experimental results in terms of number clones, \ie, the number of \alphaEquivalent goal pairs with minimum proof size of 5 lines of Ltac code, as well as the time, in seconds.
We make three time measurements: (1) under the column ``Build,'' we report the time for running Rocq on all the project files to build and generate `.vo' files; (2) under the column ``\tool Total,'' we report the end-to-end time for running \tool, including the time Rocq needs to load and type-check the files; and (2) under the column ``Analysis,'' we report the time \tool needs to parse the goals, generalize and deduplicate them, and find \alphaEquivalent ones.
Studying clones with different proof sizes is left as a future work.

In 20 out of 40 projects, \tool finds at least one clone.
The average number of goal clones per project is 27.73 (with median being 0.5).
Larger projects tend to have more instances of redundancy in them.
By inspecting the clones and their proofs, we observed 3 major types of proofs: (1) identical or near identical proofs, where the proofs are almost the same, except the variable names have changed; (2) generalized proofs, where one proof proves a more general form of the goal; (3) entirely different proofs, where new proofs have been invented in each case the goal is observed.
\begin{figure}[t!]
    \centering
    \begin{minipage}{\textwidth}
        \begin{lstlisting}[language=Coq]
            (*Goal 1:*) step_star (init ?x) ?x0 ?x1 /\ In ?x2 (messages ?x1)
            (*Proof 1:*)
                split_all.
                eassumption.
                subv s1.
                apply in_app_iff.
                right.
                apply in_eq.
            
            (*Goal 2:*) step_star (init ?x) ?x0 ?x1 /\ In ?x2 (messages ?x1)
            (*Proof 2:*)
                split_all.
                eassumption.
                subv s1.
                apply in_app_iff.
                right.
                apply in_eq.
        \end{lstlisting}
    \end{minipage}
    \caption{Duplicate proofs example: Example of two goals from Rocq project \textsf{Chapar} where \alphaEquivalent goals show up in two different theorems \texttt{KVSAlg1CauseObl.algrec\_step} and \texttt{KVSAlg1CauseObl.cause\_rec} and identical proofs have been used.}\label{fig:chapar}
\end{figure}

Figure~\ref{fig:chapar} shows an example of a pair of \alphaEquivalent goals and their identical proofs.
We do not have information about how the authors have written these proofs.
Specifically, we are not sure whether the author have copied and pasted the proofs or they have rewritten the proofs in each case.
Either way, we believe that this kind of redundancy is the least expensive one, as no new theorem or a generalization thereof needs to be created.

Figure~\ref{fig:group-theory} shows an example of the second type of redundancy, wherein one goal is in fact a generalization of the other.
In the first case, the local variable \texttt{eH} is bound to a specific value of type \texttt{U}, while in the second case, the author proves the same theorem for all values of type \texttt{U}.
In such cases, it would be possible to factor out the more general goal as an independent lemma, which could also be useful for future developments of the system.

Lastly, Figure~\ref{fig:highschool} shows an example of the third kind of redundancy, \ie, entirely different proofs have been proposed for \alphaEquivalent goals.
This could have educational value, \eg, proving same goals in different ways, or the authors have proved these goals at different times, or different people have proved the goals, and a refactoring opportunity was missed.

Another observation that we can make from Table~\ref{tab:mainTable} concerns the time measurements.
We observe that, on average, it takes 1,833.345 seconds for \tool to analyze a Rocq project in our benchmark to find clones.
We further observed that 94.71\% of this end-to-end time is from using CoqPyt/Coq-LSP to execute the Rocq files step by step, as the average time required to find clones after type checking the files is only 45.31 seconds (with median being 2.091 seconds).
This means that, compared to normal build (normal build time is reported under ``Build'' column in Table~\ref{tab:mainTable}), on average, there is a 122.4\textsf{X} increase in run time.
This significant overhead is because CoqPyt loads Rocq script files individually, rather than loading the entire project as a batch, which calls for loading and checking all the dependencies of the script file each time we load Rocq script.
We are still investigating if this issue can be resolved by only modifying CoqPyt so that its \texttt{ProofFile} class can accept a Rocq project to load, and cache the common dependencies between the files in the project and avoid loading and checking them from scratch, or if a modification to Coq-LSP is also needed.
Another way to mitigate this overhead would be through incorporating \tool as a Rocq plugin that directly retrieves the AST for goals and contexts from Rocq platform itself.
We save these engineering enhancements as a future extension of this work.
The current version of \tool tool stores partial results in file and avoids invoking Coq-LSP whenever the cached results are already available.
\begin{figure}[t!]
    \centering
    \begin{minipage}{\textwidth}
        \begin{lstlisting}[language=Coq]
            (*Goal 1:*) In U H eH -> Setsubgroup U H Gr
            (*Proof 1:*)
                intro H'.
		apply T_1_6_2 with (witness := witness); auto with sets.
		red in |- *; intros a b H'0 H'1; try assumption.
		lapply (h2 a (inv b));
                [ intro H'5; lapply H'5; [ intro H'6; generalize H'6; clear H'5 | clear H'5 ] | idtac ];
                auto with sets.
		rewrite <- (inv_involution' b); auto with sets.
		red in |- *; intros a H'0; try assumption.
		rewrite <- (G2b' (inv a)).
		apply h2; auto with sets.
		rewrite <- (G3b' witness); auto with sets.
            
            (*Goal 2:*) forall x : U, In U H x -> Setsubgroup U H Gr
            (*Proof 2:*)
                intros witness inH.
		apply T_1_6_2 with (witness := witness); trivial.
		red in |- *.
		intros a H'.
		cut (exists n : nat, inv a = phi a n).
		intro H'0; elim H'0; intros n E; rewrite E; clear H'0.
		apply positive_powers; trivial.
		cut (inv a = phi a m).
		intro H'0; rewrite H'0.
		exists m; trivial.
		symmetry in |- *.
		apply powers_repeat with (n := r); trivial.
		apply H_included_in_G; auto.
        \end{lstlisting}
    \end{minipage}
    \caption{More general proof reinvented: Example of two goals from Rocq project \textsf{GroupTheory} where \alphaEquivalent goals show up in two different theorems \texttt{T\_1\_6\_3} and \texttt{T\_1\_6\_4}. The second proof is a more general form of the first one, and the proofs seem to be reinvented for each goal.}\label{fig:group-theory}
\end{figure}

\section{Threats to Validity}\label{sec:threats}
Like all empirical studies, our findings are subject to threats to validity.
In this section we discuss some of these threats and how we mitigate them.

We had to limit our experiments to the set of Rocq projects that were compatible with Coq 8.16.0, which slashed the size of our benchmark by 86 projects.
Additionally, the Coq-LSP version shipped with Coq 8.16.0 occasionally crashed, leading to a reduced set of 40 working Rocq projects out of the 46 projects compatible with Coq 8.16.0.
Despite these cuts in the size of the benchmark, the Rocq projects are of varying sizes and domains, giving us some level of confidence about their representativeness.
Nevertheless, this reduction could introduce bias by excluding certain types of projects or proofs from the analysis, potentially affecting the representativeness of the results.
To mitigate this threat, we will make \tool open-source so that research community can study it on larger available Rocq dataset.

\tool looks for \alphaEquivalent goals in Rocq projects.
In other words, in this paper, we only consider syntactic clones.
Our results are not generalizable to the other major type of clones.
Specifically, the projects in our benchmark could contain semantic clones, \eg, closely related goals that can be generalized into a single more general goal, or even universally quantified goal pairs with non-dependent quantified variables that are different only in the order of their quantified variables.
We save this interesting topic for a future work (see ~\cref{sec:confuture}).

The basic idea behind Algorithm~\ref{alg:proof_tree_algo} is inspired from~\cite{bib:yang2019coqgym}.
Similar algorithm has been used in other works as well~\cite{bib:CompanyCoq2016,bib:roe2016coqpie}.
This style of constructing proof trees has the drawback of reporting inaccurate results when facing with compound tactics, tactics that resolves more than one goal, and focus-shifting tactics.
Some authors~\cite{bib:roe2016coqpie} have explicitly mentioned that such commands would break the algorithm, while others~\cite{bib:yang2019coqgym} have desugared or otherwise filtered out the proofs containing such tactics.
In this work, we opted for not filtering out proofs, as in our application, missing a few clone cases or reporting inaccurate proof boundaries for a goal is not harmful.

Last but not least, the primary metrics used to evaluate \tool are the number of clones detected and the time taken for analysis.
While these metrics provide significant insights into the performance of the proposed technique, they may not fully capture the practical significance or impact of the detected clones.
Additional metrics, such as the effort required to refactor or eliminate the clones will be studied in a future work.

\begin{figure}[t!]
    \centering
    \begin{minipage}{\textwidth}
        \begin{lstlisting}[language=Coq]
            (*Goal 1:*) orthogonal (vec H O) (vec H C)
            (*Proof 1:*)
                apply ortho_sym.
        	rewrite H11.
        	VReplace (vec H A) (mult_PP (-1) (vec A H)).
        	rewrite H12.
        	VReplace (mult_PP k (mult_PP (-1) (mult_PP k' (vec A B)))) (mult_PP (- (k * k')) (vec A B)).
        	auto with geo.
            
            (*Goal 2:*) orthogonal (vec H O) (vec H A)
            (*Proof 2:*)
                apply ortho_sym.
        	replace (vec H A) with (mult_PP (- k) (vec A B)).
        	auto with geo.
        	VReplace (vec H A) (mult_PP (-1) (vec A H)).
        	rewrite H8; Ringvec.
        \end{lstlisting}
    \end{minipage}
    \caption{Reinvented proofs example: Example of two goals from Rocq project \textsf{HighschoolGeometry} where \alphaEquivalent goals show up in two different theorems \texttt{intersection\_cercle\_droite} and \texttt{intersection2\_cercle\_droite}, different proofs seem to be reinvented.}\label{fig:highschool}
\end{figure}

\section{Related Work}\label{bib:related}
The works by Ghanouchi~\cite{bib:ghanouchi2023clone} and Vinan and Hupel~\cite{bib:vinan2016clone} that introduce two approaches for clone detection in Isabelle/HOL theories are closest to \tool.
Ghanouchi~\cite{bib:ghanouchi2023clone} applies fuzzy token matching to detect clones in Isabelle/HOL theory files.
This technique extends previous token-based approaches by considering near-miss clones through similarity-based token matching.
Vinan and Hupel~\cite{bib:vinan2016clone}, on the other hand, adapt the ConQAT framework, a well-known clone detection tool, to detect proof clones in Isabelle/HOL theories.
This technique extracts structured information from Isabelle document markup and performs syntactic clone detection using ConQAT’s text-based similarity analysis.
Unlike these two techniques, \tool is not intended to find clones in proof scripts themselves.
Instead, it is designed to detect goals that could be addressed once and for all.
In other words \tool leverages notions of \alphaEquivalence of goals, and proof irrelevance~\cite{bib:bertot2013interactive}, to detect redundant proof efforts.
One major benefit of \tool's clone detection approach, driven by \alphaEquivalence notion, is that it can be sound and complete, but the aforementioned similarity-based heuristics may result in false positives/negatives.

Goal clone detection aligns with \textit{proof reuse} in proof engineering.
``Large-scale proof development may involve redundant efforts that can be time-consuming.
Proof reuse addresses this by repurposing existing proofs as much as possible, minimizing the amount of redundant work that proof engineers must do''~\cite{bib:ringer2020qed}.
Existing proof reuse techniques mainly focus on proof generalization~\cite{bib:curien1995outils,bib:hasker1992generalization,bib:best2021generalizing,bib:barthe2001type,bib:ringer2019ornament,bib:johnsen2004reuse}.
To the best of our knowledge, \tool is the first technique on the topic of goal clone detection that realizes proof reuse by identifying \alphaEquivalent goals that can be extracted as independent lemmas and reused.
As such lemma extraction tools for Rocq, \eg, company-coq~\cite{bib:CompanyCoq2016} and CoqPIE~\cite{bib:roe2016coqpie}, can be used alongside \tool.

Finding \alphaEquivalent goals in Rocq proofs is directly related to the software engineering problem Type 1, \ie, exact clones, and Type 2, rename clones, clone detection.
There is a rich body of knowledge developed for clone detection for various procedural programming languages~\cite{micheline2024llm,bib:roy2007survey,bib:zakeri2023systematic,bib:white2016deep} and even Java bytecode~\cite{bib:keivanloo2012java,bib:yu2017detecting}.
Being language dependent, none of these techniques can be directly applied to goal clone detection.
Additionally, functional programming defines \alphaEquivalence, which is a theoretically sound notion of syntactic clones, \ie, Type 1 and Type 2 combined, that can be leveraged in the context of functional programming, in general, and Rocq, in particular, without resorting to potentially inaccurate heuristics proposed in the past.

\section{Conclusion and Future Work}\label{sec:confuture}
We design, implement, and evaluate a technique and a tool, named \tool, for detecting goal clones, \ie, instances of \alphaEquivalent goals that are proved multiple times within a Rocq project.
Our evaluation of \tool on 40 real-world Rocq projects reveals that half of the projects (20 out of 40) contained at least one goal clone.
The detected redundancies fell into three primary categories: (1) exact goal duplication with identical or near-identical proofs, \eg, structurally identical proofs with different variable names; (2) \alphaEquivalent goals with generalized proofs, where one proof subsumes the other; and (3) \alphaEquivalent goals with entirely different proofs.
These findings highlight a significant opportunity for improving proof maintainability and reusability in large-scale Rocq developments.

Performance analysis of \tool provided empirical evidence that it is a lightweight system, with an average runtime of 45.31 seconds per project, making it a practical tool for daily proof engineering tasks. Additionally, in the process of implementing \tool, we contributed an enhancement to CoqPyt, allowing for handling of physical-to-logical path mappings in Rocq projects.

Our results suggest that redundancy in proof engineering is a widespread issue and that automation could help mitigate wasted effort.
Moving forward, we envision extending \tool to detect semantic clones with the help of large language models, integrating it directly as a Rocq plugin for improved efficiency, and exploring automated lemma extraction to further reduce redundant proof efforts.
We are also planning to incorporate an $\eta$-equivalence detection mechanism in \tool.

\section*{Acknowledgements}
The author thanks Anonymous ECOOP 2025 Reviewers for their insightful and constructive feedback that significantly improved this paper.


\bibliographystyle{plainurl}
\bibliography{main}

\end{document}